# Chirality-induced zigzag domain wall in in-plane magnetized ultrathin films

Gong Chen[1,2,a)], MacCallum Robertson[3], Heeyoung Kwon[4], Changyeon Won[5], Andreas K. Schmid[2], and Kai Liu[1, a)]

[1] Physics Department, Georgetown University, Washington, DC 20057, USA
[2] NCEM, Molecular Foundry, Lawrence Berkeley National Laboratory, Berkeley, California, 94720 USA
[3] Physics Department, University of California, Davis, California 95616, USA
[4] Center for Spintronics, Korea Institute of Science and Technology, Seoul 02792, South Korea
[5] Department of Physics, Kyung Hee University, Seoul 02447, South Korea

[a)] Correspondence should be addressed to gchenncem@gmail.com (G.C.); Kai.Liu@georgetown.edu (K.L.)

**ABSTRACT**

The domain structure in in-plane magnetized Fe/Ni/W(110) films is investigated using spin-polarized low-energy electron microscopy. A novel transition of the domain wall shape from a zigzag-like pattern to straight is observed as a function of the film thickness, which is triggered by the transition of the domain wall type from out-of-plane chiral wall to in-plane Néel wall. The contribution of the Dzyaloshinskii-Moriya interaction to the wall energy is proposed to explain the transition of the domain wall shape, which is supported by Monte-Carlo simulations.

## I. INTRODUCTION

The formation of magnetic domain structure is a result of the interplay among competing magnetic interactions including exchange, magnetic anisotropy, dipole interaction and Dzyaloshinskii-Moriya interaction (DMI). Understanding domain configurations in various systems is fundamentally important in magnetism, and lays the foundation for device applications [1]. For example, in perpendicularly magnetized ultrathin film systems, the preferred magnetization configuration is often stripe-like patterns with opposite magnetizations in neighboring domains due to the dipole interaction [2]. The periodic width of such stripe domains reflects the balance between the long-range energy term, i.e. dipole interaction, and short-range energy terms such as exchange [3,4]. Such systems are ideal platforms for quantitative studies of these interactions [5], and for experimental control of domain patterns with tunable diploe energy via engineered magnetic coupling between multilayers [6]. The non-collinear spin structures between adjacent domains, being either helical spirals (Bloch type) or cycloidal spirals (Néel type), play important roles in the dynamic properties of domain walls, and recent discoveries of chiral domain walls in perpendicularly magnetized systems have ushered in a new era of chiral spintronics [7-16]. It is interesting to ask whether, or how, domain wall spin structure – i.e. Bloch or Néel type – may influence domain wall shapes. It is known that in perpendicularly magnetized multilayers of [Ni/Co]/Ir(111) the shape of the domains remains almost identical during the Néel to Bloch wall transition as a function of [Ni/Co]$_n$ multilayer thickness [17]. In perpendicularly magnetized systems, uniaxial in-plane anisotropy or anisotropic DMI can drive the shapes of magnetic skyrmions to become elliptical [18-20], and in systems that form a stripe domain state, the stripe-like cycloidal spirals align their boundaries normal to the easy axis of the magnetic anisotropy [21].

In in-plane magnetized films the domain structures exhibit rich varieties when the film thickness accommodates non-collinear structures along the surface normal direction [1]. Domain walls also exhibit multidimensional nature in in-plane magnetized thick films, i.e. domain wall is no longer limited to rotate in a two-dimensional plane, but rather having more complicated three-dimensional structures with both in-plane and out-of-plane components, e.g., asymmetric Bloch wall or cross-tie wall [1]. In ultrathin systems, it was widely thought that the lowest-energy domain wall spin structure is the in-plane Néel wall, i.e. the magnetization in the wall rotates within the film plane, because it minimizes the dipole energy penalty [1]. This is in sharp contrast to perpendicularly magnetized systems where the magnetization within a wall may rotate as a helical spin spiral (Bloch-type), cycloidal spin spiral (Néel-type) or a mixture of the two [22]. Recently, a novel type of chiral out-of-plane domain wall has been observed in ultrathin in-plane magnetized systems as a result of the interplay between a significant in-plane uniaxial anisotropy and a weak effective anisotropy [23], where the magnetic chirality is stabilized by the DMI [24,25]. This observation suggests that the domain wall type in ultrathin in-plane systems may be tunable, by adjusting anisotropy and DMI contributions in films with deliberately controlled layer compositions and thicknesses. The impact of in-plane domain wall spin structure on domain wall properties so far remained largely unexplored in in-plane magnetized films, e.g., the role of the domain wall type on the domain wall shape.

A key challenge in characterizing domain configurations and domain wall types is the spatial mapping of the magnetization vector, both laterally and along the depth of magnetic heterostructures. Advanced magnetic imaging techniques such as spin-polarized scanning tunneling microscopy (SP-STM) [26], scanning electron microscopy with polarization analysis (SEMPA) [27,28], spin-polarized low-energy electron microscopy (SPLEEM) [29], or photoemission electron microscopy (PEEM) [30] allow surface-sensitive detection of spin structures with nanoscale resolution. Other imaging techniques including magnetic transmission soft X-ray microscopy (MTXM) [31,32], Lorentz transmission electron microscopy (LTEM) [33,34] or electron holography [35] allow study of spin structures within magnetic nanostructures up to some thickness limits. The soft-x-ray standing wave technique developed by Fadley *et al*. has demonstrated depth-dependent imaging capabilities in probing magnetic configurations under buried surfaces and interfaces [36-38], along with other uses of x-ray photoelectron spectroscopies that reveal atomic structure and chemical information [39-42], offering exciting opportunities to understand magnetic nanomaterials.

In this paper, we have investigated magnetic domain structures in Fe/Ni/W(110) using SPLEEM. By tuning the effective magnetic anisotropy close to the spin reorientation transition through the Fe layer thickness, a transition is observed in this in-plane magnetized ultrathin film from out-of-plane domain wall spin structure to in-plane Néel wall spin structure. The shape of the magnetic domain wall qualitatively changes from zigzag-like shape to straight when the effective anisotropy shifts toward in-plane anisotropy during the Fe film growth. SPLEEM imaging of domain wall spin structures reveals that the zigzag-like wall is associated with chiral out-of-plane domain wall spin structure which lowers the energy associated with the DMI. When the domain wall evolves to an in-plane Néel wall, the DMI energy vanishes in the wall and the domain wall becomes straighter in order to minimize the wall energy. This picture is reproduced by Monte Carlo simulations, and these results provide a way to control the domain shape in ultrathin in-plane magnetized systems.

**II. EXPERIMENT**

A. Magnetic Imaging

The experiments were performed using the SPLEEM at the National Center for Electron Microscopy (NCEM) of the Lawrence Berkeley National Laboratory [29]. A GaAs-type spin polarized electron gun and spin manipulator were used to enable alignment of the electron beam spin polarization in any direction with angular resolution of ~1°. The SPLEEM allows us to map the direction of the magnetization vector by making composite images from magnetic contrast along individual in-plane and out-of-plane cartesian directions, $M_x$, $M_y$, and $M_z$ [8]. This imaging approach allows us to closely examine both the shape and chirality of the magnetic domain walls (Fig. 1). The incident electron energy is selected to be 5eV to optimize magnetic contrast. SPLEEM images were taken at room temperature, generated from the polarization-dependent reflectivities of the spin-up and spin-down electrons $I_\uparrow$ and $I_\downarrow$, respectively, where the pixel-by-pixel magnetic contrast is derived from the asymmetry of reflectivities $A = \frac{I_\uparrow - I_\downarrow}{I_\uparrow + I_\downarrow}$.

B. Sample Preparation

A W(111) substrate was cleaned by several cycles of flashing at 1950°C in 3 × 10⁻⁸ torr $O_2$ and again at the same temperature in ultrahigh vacuum with base pressure of 4 × 10⁻¹¹ torr. Fe and Ni layers were deposited on W(110) substrate at 300 K via electron beam evaporation. By monitoring oscillations in the low energy electron microscopy (LEEM) intensity associated with layer-by-layer growth, monolayer (ML) control of the film thickness was achieved. The evaporators were positioned facing the substrate at a grazing angle of 15⁰ with respect to the sample surface.

C. Data Analysis

We developed codes that locate domain wall in SPLEEM images similar to those in Fig. 2(a-d) by finding the crossover region from black to white, where the asymmetry value in the grey region is several orders of magnitude smaller compared to those in the black and white regions. Then we built up a one-pixel wide domain wall centerline, indexed these one-pixel-size points, and used the *x* and *y* coordinates to calculate the distance from each point to the adjacent point, allowing us to determine the full length of the domain wall. To produce the histograms in Fig. 2(e,f) we developed codes to determine the orientation of the domain wall tangent at each domain wall centerline pixel and measured the angle φ (see inset) between the domain wall tangent and the W[001] direction.

D. Monte Carlo Simulation

The Monte Carlo simulation was carried out on a two-dimensional model described in ref. [43], where exchange interaction, magnetic anisotropy, dipolar interaction, in-plane uniaxial magnetic anisotropy as well as the DMI are considered [23]. The dimensionless parameters $J$ (exchange), $D_\text{Dip}$ (dipole), $K_\text{eff}$ (effective anisotropy), $K_\text{u}$ (in-plane uniaxial anisotropy) and $\mathbf{D}_{ij}$ (DMI) are used for simulating domain wall spin structures. For the simulation results summarized in Fig. 3, the values $J = 1$, $D_\text{Dip}/J = 0.1$, $D_{ij}/J = 0.2$, $K_\text{u}/J = 0.05$ were assumed, and the value $K_\text{eff}/J$ is varied to capture two possible domain wall configurations. System temperature is represented by allowing spins to fluctuate according to Boltzmann statistics.

**III. RESULTS AND DISCUSSION**

SPLEEM was used to generate real space magnetic contrast images of the surface magnetization vector in ultrathin single-crystalline Fe/Ni bilayers grown on a W(110) crystal, where the Ni thickness is fixed at 15 monolayers (ML) and Fe thickness ($d_{Fe}$) ranges from 3.3 ML to 5.2 ML. The films exhibit uniaxial magnetic anisotropy with easy axis along W[001] [22,23]. A compound SPLEEM image for the $d_{Fe}$ =3.3 ML sample is shown in Fig. 1(a), where the orientation of the magnetization vector at each pixel is rendered in color according to a color wheel shown in the inset. The corresponding magnetization vector map highlights the out-of-plane domain wall between in-plane domains [Fig. 1(b)]. The spins rotate through an out-of-plane alignment from the left in-plane domain magnetized along W[00-1] to the right in-plane domain magnetized along W[001] direction. Profiles of the in-plane and out-of-plane magnetization components along the domain wall normal direction are shown in Fig. 1(c), corroborating the out-of-plane nature of the domain wall. Additionally, it is clear that the domain wall of this particular system is quite rough, in contrast to the straight walls that are seen in in-plane magnetized ultrathin systems [1,44]. The chiral character of the domain wall spin structure within the $yz$ plane is also evident. For example, a sequence of red domain (↓), white wall (⊙), cyan domain (↑), black wall (⊗), red domain (↓), white wall (⊙), and cyan domain (↑) is seen along W[001] direction ($+y$), as highlighted in the black dotted area in Fig. 1(a). Here the cycloidal-type rotation sense is fixed, i.e. domain magnetization along $y$ direction is always pointing from white wall (⊙) to black wall (⊗), consistent with previous observations in the Fe/Ni/W(110) system [23]. Note that the zigzag wall with chirality is distinct from the zigzag walls in thick films or bulk materials [1], e.g. garnet film [45], CoFeNiBSi metallic glass [46], silicon iron crystal [1,47], or iron whisker [1,48,49], where the aforementioned multidimensional domain wall structures are involved to minimize the system energies, however, without chirality [1]. The origin of the magnetic chirality in this study will be discussed later.

The thickness of the Fe overlayer is then increased to $d_{Fe}$ =5.2 ML, to induce a transition from the out-of-plane Bloch type domain wall to an in-plane Néel wall, as illustrated in Figs. 1(d,e). The domain wall shape has also changed noticeably, becoming less jagged and more straight. Fig. 1(f) corroborates the in-plane Néel wall spin texture. No clear chiral feature of the domain wall is observed in this case, which can be attributed to the vanished DMI energy in in-plane magnetized system with in-plane Néel wall [50].

To fully capture the evolution of domain shapes, we have analyzed the sequence of SPLEEM images taken during the Fe growth from $d_{Fe}$ =3.3 ML to 5.2 ML (Fig 2.a-d). The spin polarization direction of the incident electron beam is parallel to W[001] direction ($+y$), therefore the magnetization of white/black domains points in the W[001]/W[00-1] direction, as highlighted by the white arrows in Fig. 2(a). As $d_{Fe}$ increases, a rough zigzag domain wall (Fig. 2a) transforms into a smooth one (Fig. 2d), which is accompanied by a domain wall type transition from out-of-plane Bloch type to in-plane Néel type. A significant straightening of the domain wall shape is observed in the range $d_{Fe}$ =3.8 ML to 4.2 ML, suggesting that the domain wall type transition from out-of-plane to in-plane Néel likely also occurs in this $d_{Fe}$ range. To further demonstrate this evolution we have extracted histograms of angle φ (Fig. 2e,f), showing the absolute value of angle |φ| of the domain wall tangent with respect to the W[001] direction, to gauge the domain wall roughness. For $d_{Fe}$ =3.3 ML, there is a larger distribution of this angle, indicating a rougher domain wall; increasing $d_{Fe}$ to 5.2 ML, the wall becomes smoother, represented by a tighter angle distribution near φ =0°. Additionally, the length of the domain wall is plotted as a function of $d_{Fe}$ in Fig.2(g), normalized to the ideal case of a straight domain wall along $y$ direction. The gradual decrease of the normalized wall length, from ~25% larger at $d_{Fe}$ =3.3 ML to ~15% larger at $d_{Fe}$ =5.2 ML, further confirms the straightening effect.

In order to better understand the domain wall shape evolution, we have used Monte Carlo simulations to recreate the transition from the zigzag domain wall in 3.3 ML Fe/15 ML Ni/W(110) to the straight domain wall in the $d_{Fe}$ = 5.2 ML system, where the zigzag case is closer to the spin reorientation transition. Figs. 3 (a-c) show the in-plane magnetized domain configurations with different values of the effective anisotropy. Here, Fig. 3(a) exhibits an out-of-plane Bloch-type domain wall, which has a zigzag shape, whereas the in-plane Néel wall demonstrates a perfect straight shape [Fig. 3(c)]. These simulations are set up with the uniaxial anisotropy $K_u$ along the y-direction as $K_u/J$ =0.04, and the effective anisotropy $K_{eff}$ as $K_{eff}/J = 0.01$, -0.04, -0.09 for Fig.3(a-c), respectively. Here the spin reorientation transition is set as $K_{eff}/J = 0.04$ [23], and a smaller $K_{eff}/J$ value corresponds to a stronger in-plane anisotropy. In Figs.3(d-f) simplified sketches of the domain walls are shown, for a zigzag out-of-plane wall highlighting the spin rotation along $y$ direction (d) and $x$ direction (e), as well as a straight in-plane Néel wall (f). To further understand the mechanics of wall shape evolution between these three scenarios, we can perform a DMI energy analysis in a four-fold atomic lattice. This can be seen in Figs.3(g-i) where spin $S_i$ on site $i$ (light-yellow dot) is surrounded by four spins $S_j$ on sites $j$, blue arrows show the orientation of DMI vectors $D_{ij}$ between $i$ and $j$ sites [51], and orange symbols show the orientation of $S_i \times S_j$. For the case of zigzag out-of-plane wall, one could project the spin structure in such tilted wall onto $x$ (W[1-10]) and $y$ directions (W[001]) to understand the DMI energy along different directions, as shown in the dashed rectangles in (e) and (d) respectively. The DMI energy cost along $y$ direction for the decomposed spiral of such tiled wall is shown in Fig. 3 (g), where $\mathbf{D}_{ij}$ is parallel with $\mathbf{S}_i \times \mathbf{S}_j$ so the DMI energy given by $E_{DMI} = -\mathbf{D}_{ij} \cdot (\mathbf{S}_i \times \mathbf{S}_j)$ can be lowered, thus the chirality appeared along $+y$ direction as cyan domain (↑), white wall (⊙), red domain (↓) and black wall (⊗), which is consistent with the experimental observation in Fig. 1 (a). In contrast, the DMI energy vanishes in case of the decomposed spiral along the $x$ direction, as shown in Fig. 3 (e), where $\mathbf{D}_{ij}$ is perpendicular to $\mathbf{S}_i \times \mathbf{S}_j$. Therefore, in this specific domain configuration when the domain wall is parallel to the easy axis of $K_\mathrm{u}$, out-of-plane domain wall favors the zigzag shape to lower the DMI energy cost. Once the domain wall type evolves to in-plane Néel wall, $\mathbf{D}_{ij}$ will be always perpendicular to $\mathbf{S}_i \times \mathbf{S}_j$, and the DMI vanishes as well [50]. Then the domain shape prefers to be straight to minimize the domain wall energy.

To be more quantitative, the observed domain wall evolution is a result of energy minimization of the system Hamiltonian. In Fe/Ni/W(001) system, it can be written as $E = -J \sum_{<i,j>} \mathbf{S}_i \cdot \mathbf{S}_j - K_\mathrm{z} \sum_i S_{iz}^2 - K_\mathrm{u} \sum_i S_{\mathrm{u},i}^2 - D_\mathrm{Dip} \sum_{i,j} \frac{3\mathbf{S}_i \cdot (\mathbf{r}_i - \mathbf{r}_j) \mathbf{S}_j \cdot (\mathbf{r}_i - \mathbf{r}_j) - \mathbf{S}_i \cdot \mathbf{S}_j |\mathbf{r}_i - \mathbf{r}_j|^2}{|\mathbf{r}_i - \mathbf{r}_j|^5} - \sum_{<i,j>s} \mathbf{D}_{ij} \cdot (\mathbf{S}_i \times \mathbf{S}_j)$ [23], where $\mathbf{S}_i$ and $\mathbf{S}_j$ are spins located on atomic sites $i$ and $j$ in a two-dimensional plane, and $\mathbf{r}_i$ and $\mathbf{r}_j$ are the distance vectors at sites $i$ and $j$, respectively. $J$, $K_\mathrm{z}$, $K_\mathrm{u}$, $D_\mathrm{Dip}$, and $\mathbf{D}_{ij}$ correspond to exchange interaction, perpendicular anisotropy, uniaxial in-plane anisotropy, dipole interaction, and Dzyaloshinskii-Moriya interaction, respectively. From the energy minimization calculation, we found the domain wall phase boundary of the in-plane magnetized domain between the out-of-plane Bloch wall (case in Fig. 1a) and the in-plane Néel wall as $-4(1 + \sqrt{JK_\mathrm{u}/D_\mathrm{Dip}^2})$, which has an excellent agreement with the Monte Carlo simulation based phase diagram [23]. Within the phase of out-of-plane Bloch wall, the chiral feature remains the same, but the tilt angle varies in the $K_\mathrm{z} - \mathbf{D}_{ij}$ space, e.g. the tilt angle decreases when $K_\mathrm{z}$ shifts from the out-of-plane/in-plane spin reorientation transition point to the in-plane domain wall boundary, which is consistent with the experimental observation (Fig. 2). The stronger DMI also is found to induce a larger tilt. In the actual Fe/Ni/W(001) system, the DMI is estimated to be 0.53 meV/atom [23], inducing additional domain boundary tilting (30°-60° shown in Fig. 2e). We expect that the histogram of the domain wall tilt may shift its weight toward higher/lower angle with stronger/weaker DMI.

**IV. SUMMARY AND CONCLUSIONS**

In conclusion, we have investigated the domain structure in in-plane magnetized films of Fe/15ML Ni/W(110) using spin-polarized low-energy electron microscopy. A novel transition of the domain wall shape from zigzag to straight is found as Fe thickness increases from 3.3ML to 5.2ML, which is coincident with the wall type transition from out-of-plane chiral wall to in-plane Néel wall. The domain shape gradually evolves during the transition, with the total length decreasing by ~10% with respect to a perfect straight wall. This wall transition is driven by the DMI energy. In out-of-plane walls, the DMI energy favors the zigzag shape, evidenced by the presence of magnetic chirality. In in-plane Néel walls, the DMI energy vanishes, triggering the wall transition to straight walls with lowered domain wall energy. Monte-Carlo simulations have reproduced the transition of both wall type and wall shape, in excellent agreement with the experiments. Our findings suggest that the topology of domain walls may be utilized for domain pattern engineering towards novel magnetic memory and logic applications.

**ACKNOWLEDGEMENT**

This work has been supported by the NSF (DMR-2005108). Work at the Molecular Foundry was supported by the Office of Science, Office of Basic Energy Sciences, of the US Department of Energy under contract no. DE-AC02-05CH11231. We are grateful to Professor Gen Yin for helpful discussions. This contribution is dedicated to Professor Chuck Fadley, whose numerous contributions to surface/interface science and spintronics inspired the authors. Chuck was also a treasured source of wisdom and support for K.L. and his group over nearly two decades. He is dearly missed as a top-notch scientist, a beloved colleague, and a cherished mentor.

**DATA AVAILABILITY** The data that support the findings of this study are available within the article.

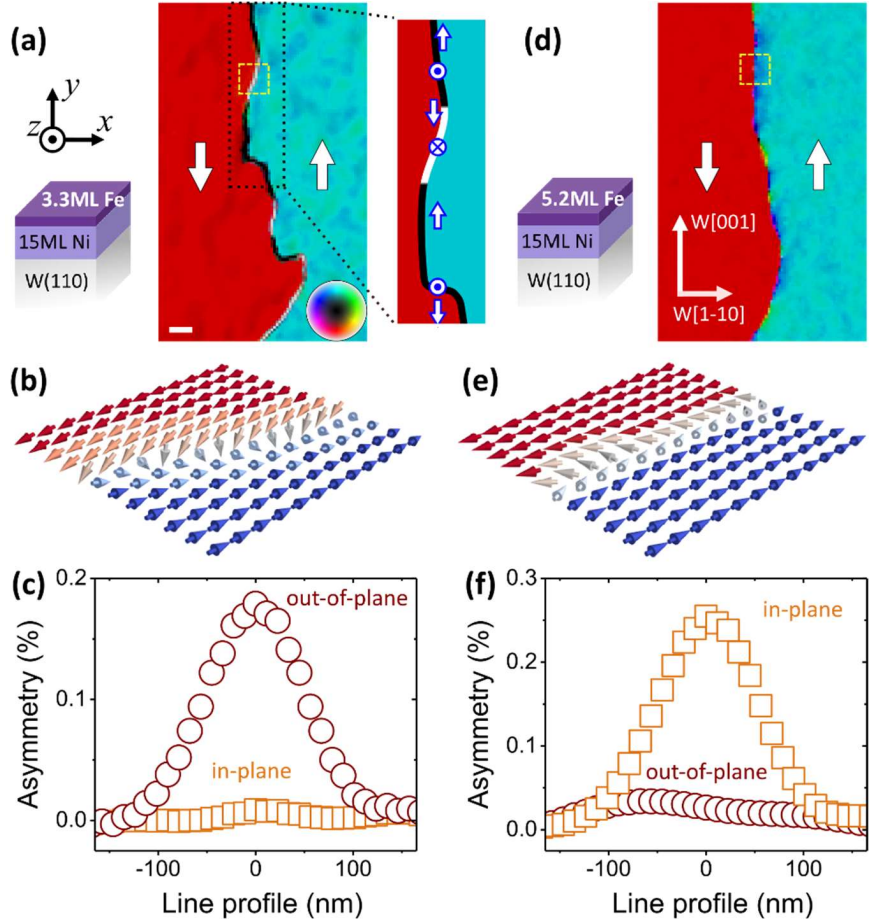

FIG. 1. Real space domain wall imaging. (a) Compound SPLEEM image of a 3.3 ML Fe/15 ML Ni/W(110) bilayer where in-plane oriented magnetization is mapped to the color wheel in the inset, with black and white representing the out-of-plane components (black being up out of page). The scale bar is 200nm. The white arrows on domains indicate the magnetization direction. The chirality is highlighted in the dotted black region, where the chiral magnetization rotation along $yz$ plane is indicated by the symbols. (b) Magnetization vectors mapped out pixel-by-pixel in a 270 x 270nm region of (a) as shown by a dashed box. (c) Averaged line profile of the out-of-plane (circles) and in-plane (squares) components along the domain wall normal for the full domain wall in (a), the asymmetry $A$ is defined in the experiment section. Corresponding plots for a 5.2 ML Fe/15 ML Ni/W(110) bilayer sample are shown in (d-f).

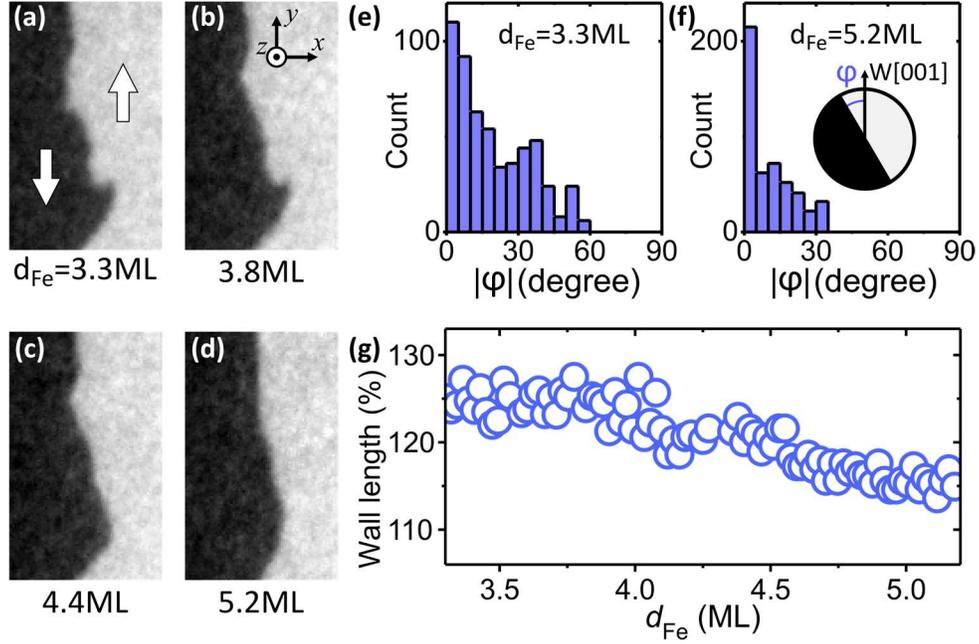

FIG. 2. Domain wall shape evolution. (a-d) Evolution of the domain wall shape in a series of averaged in-plane SPLEEM images of Fe/15 ML Ni/W(110) bilayer as a function of iron thickness $d_{Fe}$. White arrows indicate the magnetization direction in domains. Histograms showing the absolute value of the angle (φ) of the domain wall with respect to the y-axis for the (e) $d_{Fe}$ =3.3 ML and (f) $d_{Fe}$ =5.2 ML sample as a gauge of the roughness. The inset in the top right shows the angle φ with respect to the domain wall. (g) Plot of domain wall length as a function of $d_{Fe}$ taken from the non-averaged raw in-plane SPLEEM images in (a-d). The wall lengths are normalized to the length of a completely straight domain wall across the imaging region. Panels (a-d) are adapted with permission from G. Chen, S. P. Kang, C. Ophus, A. T. N'Diaye, H. Y. Kwon, R. T. Qiu, C. Won, K. Liu, Y. Wu, and A. K. Schmid, Out-of-plane chiral domain wall spin-structures in ultrathin in-plane magnets, Nat. Commun. 8, 15302, (2017); licensed under a Creative Commons Attribution 4.0 International (CC BY 4.0) license.

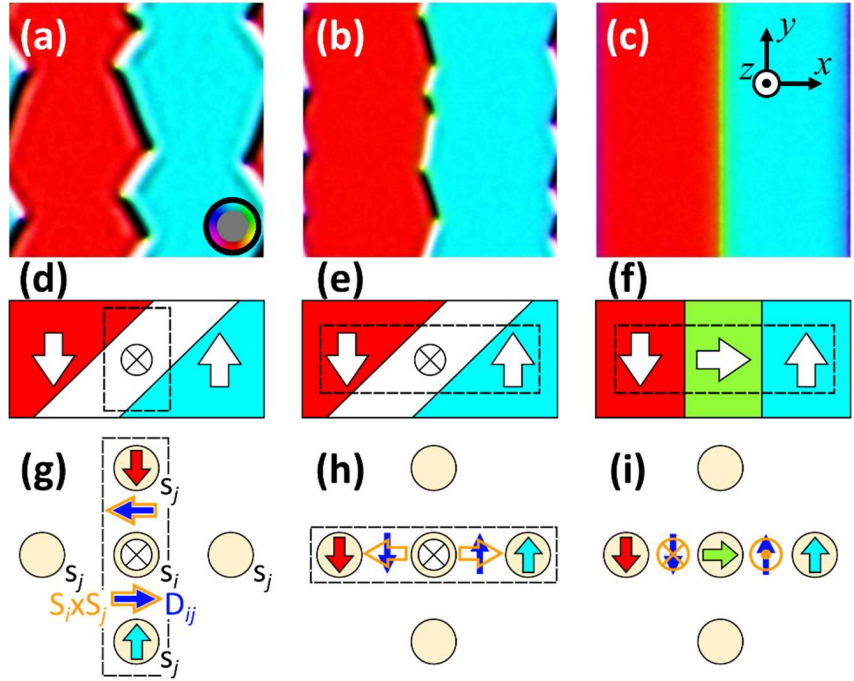

FIG. 3. The mechanism of the wall-type induced domain wall evolution. (a-c) Monte Carlo simulation of the magnetic domain configurations with various magnetic parameters, the uniaxial anisotropy $K_u$ along the $y$ direction is set as $K_u/J =0.04$, and the effective anisotropy $K_{eff}$ is set as $K_{eff}/J =0.01, -0.04, -0.09$ in panel a to c, respectively. Here the in-plane to out-of-plane spin reorientation transition occurs at $K_{eff}/J =0.04$. (d-f) Simplified sketch of domain configurations near domain boundaries, arrows represent the magnetization direction in domains or domain walls, out-of-plane wall with a tilted domain wall in panel (d), out-of-plane wall with a vertical wall in panel (e), in-plane Néel wall in a vertical domain wall in panel (f). (g-i) DMI energy analysis in a four-fold atomic lattice, where spin $S_i$ on $i$ site (light-yellow dot) surrounded by four spins $S_j$ on $j$ sites. Solid blue arrows show the orientation of DMI vectors $D_{ij}$ between $i$ site and $j$ site. Hollow orange symbols show the orientation of $S_i \times S_j$. The DMI energy $E_{DMI} = -D_{ij} \cdot (S_i \times S_j)$ is non-zero and thus the overall energy of the system lowered when $D_{ij}$ is parallel to $S_i \times S_j$ in panel g, and the DMI energy vanishes in panel g and h.